\documentclass[12pt,a4paper]{article}

\usepackage{graphics}
\usepackage{latexsym}
\usepackage{amsmath}
\usepackage{amssymb}
\usepackage{slashed}
\usepackage[usenames]{color}

\title{Understanding the Formation of Galaxies with Warm Dark Matter}
\author{Bruce Hoeneisen}
\date{\small{
Universidad San Francisco de Quito, Quito, Ecuador \\
Email: bhoeneisen@usfq.edu.ec \\
8 October 2023}
}

\begin{document}
\maketitle

\begin{abstract}
\noindent
The formation of galaxies with warm dark matter
is approximately adiabatic. The cold dark matter
limit is singular and requires relaxation.
In these lecture notes we develop, step-by-step,
the physics of galaxies	with warm dark matter,
and their formation. The theory	is validated
with observed spiral galaxy rotation curves.
These observations constrain the properties of
the dark matter particles.

\end{abstract}


\section{Introduction}
\label{introduction}

The formation of galaxies is qualitatively different if
dark matter is warm instead of cold. The cold dark matter
limit is singular.
It turns out that understanding galaxies, and the formation
of galaxies, is \textit{less difficult} if dark matter
is warm. 
So, we add to the cold dark matter
$\Lambda$CDM cosmology one more parameter, namely the
temperature-to-mass ratio of dark matter, and let
observations decide whether dark matter is warm or cold.
These lecture notes do bring new understanding of dark matter,
and constrain the properties of the dark matter particles.

\section{Warm dark matter}
\label{wdm}

Let us consider warm dark matter as a non-relativistic classical noble gas
of particles of mass $m$, density $\rho$ and temperature $T$.
By ``classical" we mean that the velocity distribution of the non-relativistic
particles of dark matter in the early universe
is assumed to be the Maxwell distribution, i.e. is not degenerate.
For a discussion on how dark matter might have acquired the Maxwell
distribution of velocities, see \cite{Pfenniger}.
By ``noble" we mean that collisions, if any, do not excite internal
states of the particles. 
Recall that for a non-relativistic gas, $\frac{1}{2} m \left< v^2 \right> = \frac{3}{2} kT$.
$\sqrt{\left< v^2 \right>}$ is the root-mean-square thermal velocity of the 
dark matter particles.
Recall that for adiabatic expansion of a noble gas,
$T V^{\gamma - 1} = T V^{2/3} = $ constant (here $V$ is the volume), so 
\begin{equation}
\sqrt{\left< v^2 \right>} \rho^{-1/3} = \textrm{constant}
\label{ai}
\end{equation}
is an adiabatic invariant.
We will review these concepts in more detail in the next lecture.
We want to stress that equation (\ref{ai}) is valid for a collisionless or collisional gas,
and is valid whether, or not, the particles are bouncing off the walls
of an expanding box of volume $V$.

To verify these statements,
let us now consider a free particle in a homogeneous universe with
expansion parameter $a(t)$, normalized to $a(t_0) = 1$ at the present
time $t_0$. The particle has velocity $v$ at the space point $\mathbf{r}$.
In time $dt$ the particle advances $v dt$ and arrives at the space point
$\mathbf{r}'$. Due to the expansion of the universe,
$\mathbf{r}'$ moves away from $\mathbf{r}$ with velocity
\begin{equation}
H v dt = \frac{1}{a} \frac{da}{dt} v dt = \frac{da}{a} v.
\end{equation}
So the velocity of the free particle relative to $\mathbf{r}'$,
when it coincides with $\mathbf{r}'$, is reduced by
$-dv = (da/a) v$, so
\begin{equation}
v a = \textrm{constant}.
\end{equation}
Note that this expression is in agreement with equation (\ref{ai}).
As the universe expands, velocities decrease in proportion to $a^{-1}$,
and temperatures decrease in proportion to $a^{-2}$.

It is not necessary to invoke General Relativity. The preceding
arguments are valid also in the non-relativistic physics of Newton.
During galaxy formation, the dark matter particles have a velocity 
field (to be shown in Figure \ref{galaxy_h_b_12} below) 
in addition to the thermal velocity. 
Whenever the velocity field diverges, dark matter cools 
as described in the preceding paragraph.

The density $\rho(a)$ of matter is proportional to $a^{-3}$.
We define the ``adiabatic invariant" of warm dark matter as
the comoving root-mean-square thermal velocity of the dark matter particles:
\begin{equation}
v_{h\textrm{rms}}(1) \equiv v_{h\textrm{rms}}(a) a = 
v_{h\textrm{rms}}(a) \left( \frac{\Omega_c \rho_\textrm{crit}}{\rho_h(a)} \right)^{1/3}. 
\label{v_hrms}
\end{equation}
$\Omega_c \rho_\textrm{crit}$ is the mean dark matter density of the universe
at the present time
(throughout we use the notation and the values of parameters of \cite{PDG2020}).
The sub-index $h$ stands for the dark matter halo.
We will use the sub-index $b$ for ``baryons", mostly hydrogen and helium.
These sub-indices will be used only when needed.

Consider a free observer in a density peak in the early universe.
This observer ``sees" warm dark matter expand adiabatically, 
reach maximum expansion, and then contract adiabatically into the
core of a galaxy.
The adiabatic invariant (\ref{v_hrms}) in the early universe remains constant 
in the core of the galaxy throughout the formation of the galaxy.
This statement is non-trivial, so we will study it in detail in
the following lectures, and finally will validate it with observations.

To the cold dark matter cosmology $\Lambda$CDM, that has six parameters,
we add one more parameter, namely the adiabatic invariant $v_{h\textrm{rms}}(1)$,
and obtain the warm dark matter cosmology $\Lambda$WDM.
As we shall learn in the following lectures, we are able to measure
$v_{h\textrm{rms}}$ and $\rho_h$ in the core of spiral galaxies, and will
therefore be able to obtain $v_{h\textrm{rms}}(1)$, and then decide whether
dark matter is warm or cold.

\section{The exponential isothermal atmosphere}

This lecture is included to remind the reader of results
we will be using later on. (For more background, I recommend
studying the awe inspiring Feynman Lectures on ``The exponential
atmosphere".)

James Clerk Maxwell presented several clever arguments to obtain the
distribution of velocities of the particles in a gas in thermal equilibrium.
The number of particles per unit phase space volume
$d^3\mathbf{r} d^3\mathbf{p} \equiv dx dy dz dp_x dp_y dp_z$ is 
proportional to $\exp{\left[ -p^2/(2mkT) \right]}$:
\begin{equation}
\frac{dn}{d^3\mathbf{r} d^3\mathbf{p}} \propto \exp{\left[ -p^2/(2mkT) \right]},
\end{equation}
where $\mathbf{p} = m\mathbf{v}$ is the particle momentum.
The momenta $\mathbf{p}$ are assumed isotropic (later on we will lift this assumption
when needed).
This Maxwell distribution has been validated experimentally
(which settles the issues of the clever arguments), and is our point of departure
in these lectures.
Using the definite integrals
\begin{equation}
\int_0^\infty e^{-x^2} x^2 dx = \frac{\sqrt{\pi}}{4}, \qquad \int_0^\infty e^{-x^2} x^4 dx = \frac{3 \sqrt{\pi}}{8}, 
\end{equation}
the following well known results are readily obtained:
\begin{equation}
\frac{1}{2} m \left< v^2 \right> = \frac{3}{2} kT, 
\qquad P = \frac{\rho}{m} kT \qquad \textrm{with} \qquad \rho \equiv \frac{N}{V} m,
\end{equation}
where the pressure $P$ is defined as twice the momenta $p_z$ of particles 
with $p_z > 0$ traversing unit area in unit time.

Let $N/V$ be the number of particles per unit volume of the gas.
Then the normalized Maxwell distribution is
\begin{equation}
\frac{dn}{d^3\mathbf{r} d^3\mathbf{p}} = \frac{N}{V} \frac{1}{(2 \pi m k T)^{3/2}} 
\exp{\left[ \frac{-p^2}{2mkT} \right]}.
\label{MB}
\end{equation}
The number of particles with momenta in the interval $p_z$ to $p_z + dp_z$,
traversing unit area in unit time, obtained from (\ref{MB}), is
\begin{equation}
dF = v_z dp_z \int \frac{N}{V} \frac{1}{(2 \pi m k T)^{3/2}} \exp{\left[ \frac{-p^2}{2mkT} \right]} dp_x dp_y.
\label{dF}
\end{equation} 

Now consider gas in equilibrium in a column in a uniform gravitational 
field $\mathbf{g} = g_z \mathbf{e}_z$ with $g_z < 0$.
The potential energy of a particle at altitude $z = h$ is $\Phi = m (-g_z) h$.
We wish to obtain the distribution of momenta at altitude $h$.
We consider particles with $p_z > 0$ and $p_z^2/(2 m) > \Phi$,
with momenta in the interval $p_z$ to $p_z + dp_z$.
These particles reach altitude $h$ 
with momenta in the interval $p'_z$ to $p'_z + dp'_z$, where
\begin{equation}
\frac{p_z^2}{2m} = \frac{p'^2_z}{2m} + \Phi, \qquad \textrm{and} \qquad
\frac{(p_z + dp_z)^2}{2m} = \frac{(p'_z + dp'_z)^2}{2m} + \Phi.
\end{equation}
Subtracting these two equations, and keeping first order terms, obtains
\begin{equation}
v_z dp_z = v'_z dp'_z.
\end{equation}
Multiplying by $dt$ obtains
\begin{equation}
dz dp_z = dz' dp'_z.
\end{equation}
This equation is Liouville's theorem in one dimension:
the volume of phase space occupied by particles moving in a conservative
field is a constant of the motion.

In equilibrium, the flux $dF'$  of particles at altitude $h$ with 
momenta in the interval $p'_z$ to $p'_z + dp'_z$
is the same as the flux $dF$ of particles at altitude $0$ with
momenta in the interval $p_z$ to $p_z + dp_z$, since these are
the same particles. Therefore, expressing $dF$ in terms of variables at altitude $h$,
we obtain:
\begin{equation}
dF = dF' = v'_z dp'_z \int \left( \frac{N}{V} \right)' \frac{1}{(2 \pi m k T)^{3/2}} 
\exp{\left[ \frac{-p'^2}{2mkT} \right]} dp'_x dp'_y.
\label{dF'}
\end{equation}
where
\begin{equation}
\left( \frac{N}{V} \right)' = \frac{N}{V} \exp{\left[ \frac{-\Phi}{kT} \right]}
\end{equation}
So, the density varies with altitude as
\begin{equation}
\rho(h) = \rho(0) \exp{\left[ \frac{-\Phi}{kT} \right]}.
\label{rho}
\end{equation}
Comparing (\ref{dF'}) with (\ref{dF}), we obtain the distribution of 
particles at any altitude:
\begin{equation}
\frac{dn}{d^3\mathbf{r} d^3\mathbf{p}} \propto \exp{\left[ -\frac{E}{kT} \right]},
\label{dnh}
\end{equation}
where the energy of a particle is
\begin{equation}
E = \frac{p^2}{2m} + \Phi(\mathbf{r}).
\label{E}
\end{equation}
Equation (\ref{dnh}) is the Boltzmann distribution.
The proportionality constant in (\ref{dnh}) is independent of altitude.
Note that $T$ in (\ref{dnh}) is the same $T$ as in (\ref{MB}), so the
column of gas is isothermal!
These equations are valid for general potential energies $\Phi$, whether, or
not, proportional to $h$.

An amazing result of these calculations is that the root-mean-square
velocity $\sqrt{ \left< v^2 \right> }$ of the particles of the gas is independent of altitude!
This is because only the more energetic particles reach a higher altitude.
Another amazing result is that the column of gas is isothermal
because of the Maxwell distribution of momenta, not because of thermal
contact, or not, with the walls (if any) of the gas column.

We have not considered particle collisions. It turns out that results are unchanged because
collisions conserve energy.

If the particles are collisionless, the thermal velocities may not be isotropic,
in which case we will consider separately each component of the thermal velocity.

One more equation that we will be using in the sequel is
\begin{equation}
\frac{dP}{dz} = \frac{m g_z}{kT} P = \rho g_z,
\label{P}
\end{equation}
in agreement with (\ref{rho}).
This equation expresses that the difference of pressure $P(z) - P(z +dz)$
supports the weight of the gas between $z$ and $z + dz$.
This result may again be surprising since there are no membranes at
$z$ and $z + dz$ on which the particles can bounce off.
Equation (\ref{P}) expresses conservation of momentum.

\section{The cored isothermal sphere}
\label{is}

Let us repeat the arguments of the preceding lecture, but this time
we consider a self-gravitating gas with spherical symmetry.
The equations to be solved are
\begin{equation}
\nabla \cdot \mathbf{g} = \frac{1}{r^2} \frac{d}{dr} (r^2 g_r) = -4 \pi G \rho, \qquad
\nabla P = \frac{dP}{dr} \mathbf{e}_r = \rho \mathbf{g}, \qquad
P = \left< v^2_{r} \right> \rho.
\label{is_eq}
\end{equation}
The rotation velocity $V(r)$ of a test particle in a circular orbit of radius $r$ is given by 
\begin{equation}
-g_r(r) = \frac{V(r)^2}{r}.
\end{equation}
First we seek a particular solution of the form $\rho(r) \propto r^n$,
with $\left< v^2_{r} \right>$ independent of $r$, i.e. we limit the scope of
the present lectures to the isothermal case.
There is a single solution of this form (with $n = -2$):
\begin{equation}
\rho(r) = \frac{kT}{2 \pi G m r^2}, \qquad M(r) = \frac{2 k T}{Gm} r, \qquad
V = \sqrt{\frac{2 k T}{m}} = \sqrt{ 2 \left< v_r^2 \right>}.
\label{is_sol}
\end{equation}
The potential energy difference	between	particles at $r'$ and $r$
is \\
$\Phi = 2kT \ln{(r'/r)}$, so (\ref{rho}) is valid.
There are $dn$ particles in phase space	volume $4 \pi r^2 dr dp_r$,
that will later occupy phase space volume $4 \pi r'^2 dr' dp'_r$.
Then
\begin{eqnarray}
dF & \equiv & \frac{dn}{4 \pi r^2 dt} = \left( \frac{dn}{d^3\mathbf{r} d^3\mathbf{p}} \right) v_r dp_r, \\
dF' & \equiv & \frac{dn}{4 \pi r'^2 dt'} = \left( \frac{dn}{d^3\mathbf{r} d^3\mathbf{p}} \right)' v'_r dp'_r,
\end{eqnarray}
so $dF'	= (r/r')^2 dF$.
The total mass of the halo can be defined as $M(r_\textrm{max})$ with
$\rho(r_\textrm{max}) =	\Omega_c \rho_\textrm{crit}$. Then
$M(r_\textrm{max}) \propto T^{3/2} \propto V^3$, which is the Tully-Fisher
relation if $M(r_\textrm{max})$	is proportional	to the absolute	luminosity.

Equations (\ref{is_eq}) can be solved numerically in radial steps $dr$,
starting from $r_\textrm{min}$.
To start the numerical integration we need to provide two
boundary conditions, e.g. $g_r(r_\textrm{min})$, and $\rho(r_\textrm{min})$.
For solutions with no black hole at $r=0$ we can take
$g_r(r_\textrm{min}) \approx -G 4 \pi r_\textrm{min} \rho(r_\textrm{min})/3$.
For these isothermal spheres with a core the solution for
$r \gg r_c$ is (\ref{is_sol}), while the solution for
$r \ll r_c$ is 
\begin{equation}
\rho(r \ll r_c) = \rho_0, \qquad V(r \ll r_c) = \sqrt{\frac{4}{3} \pi G \rho_0} r.
\end{equation}
The two asymptotes meet at a core radius
\begin{equation}
r_c = \sqrt{ \frac{3 \left< v_r^2 \right>}{2 \pi G \rho_0}}.
\end{equation}

Since the cored isothermal sphere has formed
from a density perturbation in the early universe, 
the two parameters $\rho_0$ and $\left< v_r^2 \right>$ are related by
the adiabatic invariant (\ref{v_hrms}):
\begin{equation}
v_{h\textrm{rms}}(1) = 
\sqrt{3 \left< v_r^2 \right>} \left( \frac{\Omega_c \rho_\textrm{crit}}{\rho_0} \right)^{1/3}.
\label{v_hrms_2}
\end{equation}
Therefore the cored isothermal sphere is defined by a single 
independent parameter, $\rho_0$ or $\left< v_r^2 \right>$.
Equation (\ref{v_hrms_2}) assumes isotropic velocities in the core, which is
valid due to the spherical symmetry.
Further justification of (\ref{v_hrms_2}), and observational validation, 
will be given in following lectures.

We note that a measurement of $V(r \gg r_c)$ obtains $\sqrt{\left< v_r^2 \right>}$,
and a measurement of $dV(r \ll r_c) / dr$ obtains $\rho_0$, and together
they obtain the adiabatic invariant $v_{h\textrm{rms}}(1)$.

\section{An example}

\begin{figure}
\begin{center}
%\vspace*{-4.5cm}
\scalebox{0.5}
%{\includegraphics{Fig_UGCA442_020819.eps}}
{\includegraphics{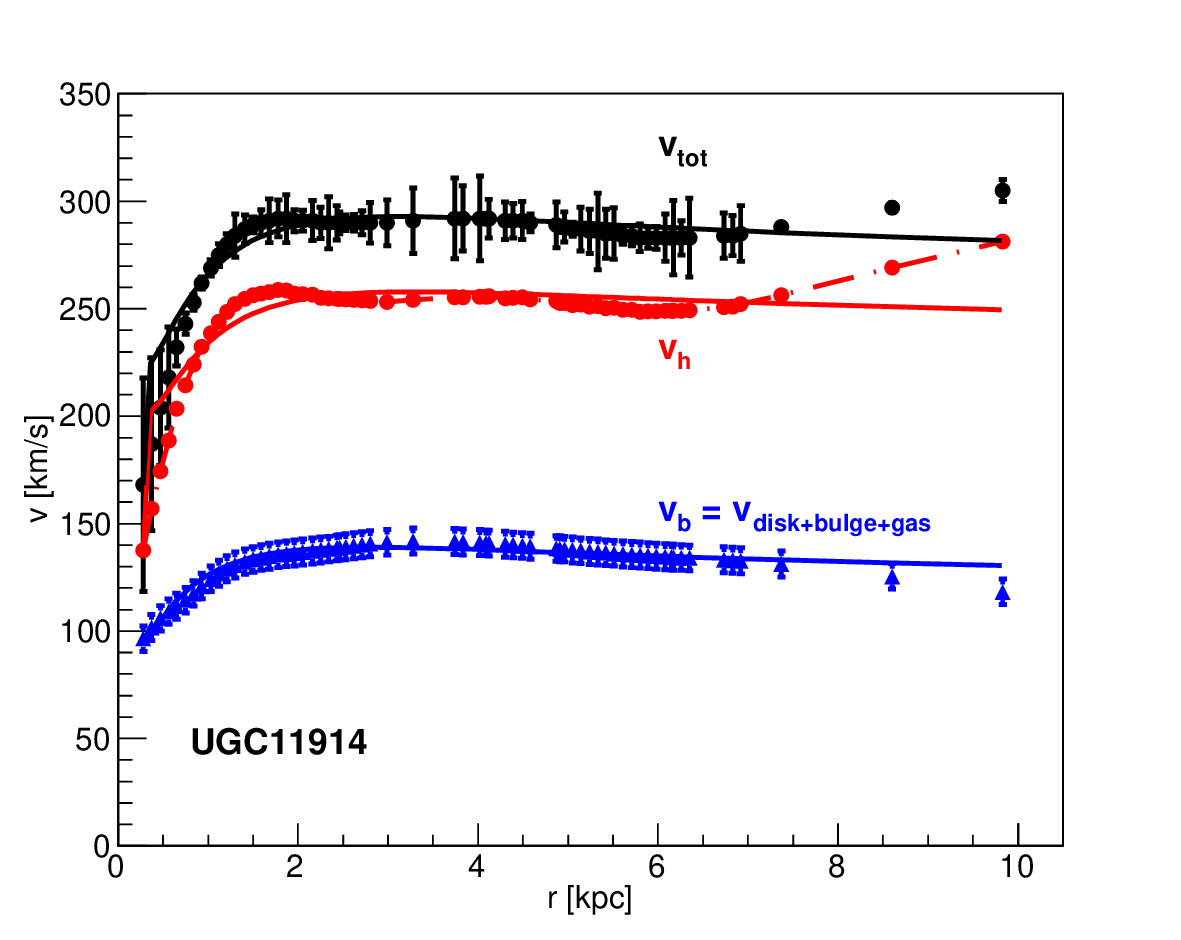}}
\scalebox{0.5}
%{\includegraphics{Fig_UGCA442_rho_020819.eps}}
{\includegraphics{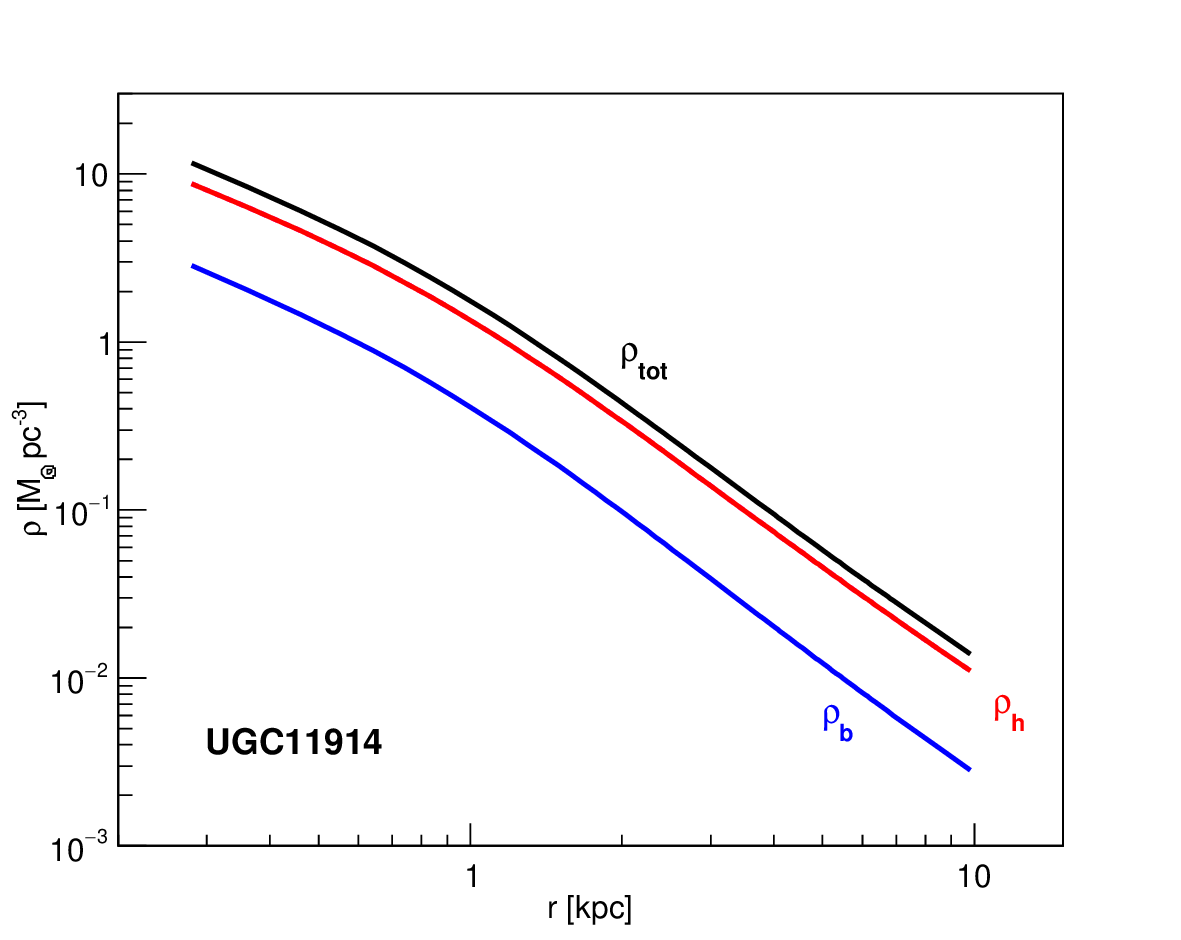}}
%\vspace*{0.7cm}
\caption{
Top: Observed rotation curve $V_\textrm{tot}(r)$ (dots)
and the baryon contribution $V_b(r)$ (triangles) of galaxy UGC11914 \cite{SPARC}.
The solid lines are obtained by numerical integration
as explained in the text.
Bottom:
Mass densities of baryons and dark matter
obtained by the numerical integration.
}
%\label{UGCA442}
\label{UGC11914}
%https://www.scirp.org/pdf/IJAA_2019092915434576.pdf
%/home/bruce1/vaccuum/SPARC/LTG/ijaa7/tmp/tmp1/adiabatic_invariant_3_final.tex
%iso_galaxies_0810_2100_SPARC_LITTLE_THINGS.C_bck011023_UGC11914_2
\end{center}
\end{figure}

As an example with extremely large $\rho_0$, let us consider the spiral galaxy UGC11914.
The rotation curves of this galaxy, observed by the SPARC collaboration \cite{SPARC},
are presented in Figure \ref{UGC11914}.
We fit these rotation curves by solving the equations:
\begin{eqnarray}
\nabla \cdot \mathbf{g}_b & = & -4 \pi G \rho_b, \qquad
\nabla \cdot \mathbf{g}_h = -4 \pi G \rho_h, \label{g} \\
\mathbf{g} & = & \mathbf{g}_b + \mathbf{g}_h, \qquad
g_b \equiv -\frac{v_b^2}{r}, \qquad g_h \equiv -\frac{v_h^2}{r},
\qquad V^2 = v_b^2 + v_h^2, \label{v} \\
\nabla P_b & = & \rho_b \left( \mathbf{g} + \kappa_b \frac{V^2}{r} \hat{\mathbf{e}}_r \right), \qquad
\nabla P_h = \rho_h \left( \mathbf{g} + \kappa_h\frac{V^2}{r} \hat{\mathbf{e}}_r \right),  \label{P3} \\
P_b & = & \left< v_{rb}^2 \right> \rho_b, \qquad \textrm{and} \qquad
P_h = \left< v_{rh}^2 \right> \rho_h. \label{P2}
%/home/bruce1/vaccuum/dark_matter_v6.tex
\end{eqnarray}
We have included $\kappa_b$ and $\kappa_h$ to account for rotation.
These parameters are quite uncertain, tho a rough estimate is $\kappa_b \approx 0.98$ and 
$\kappa_h \approx 0.15$ \cite{part1}.
We can eliminate $\kappa_b$ and $\kappa_h$ from the numerical integration by replacing 
$\left< v_{rb}^2 \right> \rightarrow \left< v_{rb}^2 \right> / (1 - \kappa_b)$,
and $\left< v_{rh}^2 \right> \rightarrow \left< v_{rh}^2 \right> / (1 - \kappa_h)$.
We start the numerical integration at the first measured point at $r_\textrm{min}$,
and end at the last measured point at $r_\textrm{max}$, in steps $dr$.
To start the numerical integration we need six boundary conditions:
$\left< v_{rb}^2 \right> / (1 - \kappa_b)$, $\left< v_{rh}^2 \right> / (1 - \kappa_h)$,
$\rho_b(r_\textrm{min})$, $\rho_h(r_\textrm{min})$, 
$M_{bBH}$, and $M_{hBH}$, where
\begin{eqnarray}
g_b(r_\textrm{min}) & = & - G (4/3) \pi r_\textrm{min} \rho_b(r_\textrm{min}) - G M_{bBH}/r_\textrm{min}^2, \\
g_h(r_\textrm{min}) & = & - G (4/3) \pi r_\textrm{min} \rho_h(r_\textrm{min}) - G M_{hBH}/r_\textrm{min}^2,
\end{eqnarray}
to include the possibility of a black hole at the center.
Good fits are obtained assuming $\left< v_{rb}^2 \right> / (1 - \kappa_b)$ and 
$\left< v_{rh}^2 \right> / (1 - \kappa_h)$ are independent of $r$.
We vary the six boundary conditions to minimize the $\chi^2$ between the observed and
calculated rotation velocities, and obtain:
\begin{eqnarray}
\sqrt{ \frac{\left< v_{rb}^2 \right>}{1 - \kappa_b}} & = & 193.9 \pm 1.7 \textrm{ km/s}, \qquad
\sqrt{ \frac{\left< v_{rh}^2 \right>}{1 - \kappa_h}} = 197.5 \pm 0.8 \textrm{ km/s}, \\
\rho_b(r_\textrm{min}) & = & 2.86 \pm 0.22 \textrm{ M}_\odot/\textrm{pc}^3, \qquad
\rho_h(r_\textrm{min}) = 8.75 \pm 0.29 \textrm{ M}_\odot/\textrm{pc}^3, \\
M_{bBH} & = & (2.85 \pm 0.95) \times 10^8 \textrm{ M}_\odot, 
M_{hBH} = (18.6 \pm 2.0) \times 10^8 \textrm{ M}_\odot.
\end {eqnarray}
Uncertainties are statistical from the fit.
The core radius is $0.70$ kpc.
The core dark matter density 
$\rho_h(r_\textrm{min})$ is $2.6 \times 10^8$ times the mean
dark matter density $\Omega_c \rho_\textrm{crit}$.
We note that this galaxy indeed has a black hole at its center.
From $\left< v_{rh}^2 \right>$ and $\rho_h(r_\textrm{min})$ we obtain the
adiabatic invariant
\begin{equation}
v_{h\textrm{rms}}(1) = \sqrt{(1 - \kappa_h)} \cdot (535 \pm 8) \textrm{ m/s}.
\label{v_hrms_UGC11914}
\end{equation}

Let us mention that similar results are obtained from galaxies spanning $3.5$
orders of magnitude in absolute luminosity \cite{dwarf}, 
validating that the adiabatic invariant in the core of galaxies is
conserved, confirming that $v_{h\textrm{rms}}(1)$ 
is of cosmological origin, and that dark matter is indeed warm!

If we add an $r$-dependence to $\left< v_{rh}^2 \right>$ we generally
obtain a higher $\chi^2$ of the fit, so indeed, within uncertainties, the dark matter halo is
isothermal, at least out to the observed rotation curves,
indicating that the particles have the Maxwell-Boltzmann momentum distribution.
We note that $\sqrt{\left< v_{rb}^2 \right>/(1 - \kappa_b)}$ is of the same
order of magnitude as $\sqrt{\left< v_{rh}^2 \right>/(1 - \kappa_h)}$, because
dark matter and baryons fall into the same potential well.
Since thermal equilibrium implies 
$\frac{1}{2} m_p \left< v_{rb}^2 \right> = \frac{1}{2} m_h \left< v_{rh}^2 \right>$,
we conclude that, unless $m_p (1 - \kappa_b) \approx m_h (1 - \kappa_h)$, 
dark matter is not in thermal
equilibrium with baryons (mostly hydrogen atoms).
So, on galactic scales, we can neglect dark matter-baryon interactions.
If dark matter feels only the gravitational interaction, it can be shown
that deviations of its trajectory, or interchange of energy with baryons, can be
neglected on galactic scales.
We also note that the ratio of dark matter to baryon densities in the galaxy is of
the order of the universe average.

The galaxy density (\ref{is_sol}) reaches $\Omega_c \rho_\textrm{crit}$ at
\begin{equation}
r_\textrm{max} = \left( \frac{\left< v_{rh}^2 \right>}
{2 \pi G \Omega_c \rho_\textrm{crit}} \right)^{1/2} \approx 6.6 \textrm{ Mpc}.
\end{equation}
The age of the universe at, say, redshift $z = 6$, is $3 \times 10^{16}$ s.
In this time a particle with constant velocity $198$ km/s reaches 0.2 Mpc,
less than $r_\textrm{max}$, and much farther than the last observed
rotation velocity.

From (\ref{v_hrms_UGC11914}), neglecting $\kappa_h$, we estimate that dark matter 
becomes non-relativistic at expansion parameter
\begin{equation}
a_{h\textrm{NR}} = \frac{v_{h\textrm{rms}}(1)}{c} \approx 1.8 \times 10^{-6},
\end{equation} 
i.e. after $e^+ e^-$ annihilation, and while the universe is still dominated
by radiation.

Now let us do the following back-of-the-envelope approximate calculations.
An ultra-relativistic gas with zero chemical potential has a
number density of particles $n(T) = 0.1218 \cdot (k T / (\hbar c))^3 (N_b + 3 N_f / 4)$ \cite{PDG2020}.
At expansion parameter $a_{h\textrm{NR}}$ the temperature 
is $T_h(a_{h\textrm{NR}}) \approx m_h c^2/(3 k)$.
The present number density is 
$n(1) = n(a_{h\textrm{NR}}) a_{h\textrm{NR}}^3 = \Omega_c \rho_\textrm{crit} / m_h$,
if dark matter particles do not decay or annihilate
at $\approx a_{h\textrm{NR}}$, i.e. if there is no ``freeze-out".
From these equations we obtain
\begin{eqnarray}
m_h & \approx & \left( \frac{\Omega_c \rho_\textrm{crit} (3 \hbar)^3}
{0.1218 \cdot v^3_{h\textrm{rms}}(1)} \right)^{1/4} \left( N_b + 3 N_f/4 \right)^{-1/4}, \qquad \textrm{or} \\
\frac{m_h c^2}{e}  & \approx & 107.3 \textrm{ eV} \left( \frac{760 \textrm{ m/s}}{v_{h\textrm{rms}}(1)} \right)^{3/4} 
          \left( N_b + 3 N_f/4 \right)^{-1/4}.
\label{mh}
\end{eqnarray}
For scalar dark matter, i.e. $N_b = 1$ and $N_f = 0$,
and neglecting $\kappa_h$, 
we estimate the mass of the dark matter particles from (\ref{v_hrms_UGC11914}):
$m_h c^2/e \approx 140$ eV.
At expansion parameter $a_{h\textrm{NR}}$ the photon temperature 
is $T_\gamma(a_{h\textrm{NR}}) = T_0/a_{h\textrm{NR}}$.
From the preceding equations we obtain
\begin{equation}
\frac{ T_h(a_{h\textrm{NR}})}{T_\gamma(a_{h\textrm{NR}})} \approx 
0.386 \left( \frac{v_{h\textrm{rms}}(1)}{760 \textrm{ m/s}} \right)^{1/4} 
\left( N_b + 3 N_f/4 \right)^{-1/4},
\label{Th_Tg}
\end{equation}
or, for our example, $T_h(a_{h\textrm{NR}}) / T_\gamma(a_{h\textrm{NR}}) \approx 0.354$.
That this ratio is of order 1, given that the dark matter mass is
uncertain over 89 orders of magnitude \cite{PDG2020}, is surely
telling us something! Furthermore,
note that dark matter is sufficiently cooler than photons to (marginally)
evade the ``thermal relic" mass limits obtained from the Lyman-$\alpha$ forest \cite{comparing},
and sufficiently cool to not spoil the success of Big-Bang Nucleosynthesis \cite{part1}.
Finally, within experimental uncertainties, dark matter is in
thermal and diffusive equilibrium with the Standard Model sector
at $T$ somewhere between the top quark mass $m_t$ and the
temperature $T_C$ of the deconfinement-confinement transition 
from quarks to hadrons (decoupling at a lower temperature 
compromises the agreement with Big Bang Nucleosynthesis \cite{part1}). 
For a proper treatment of the preceding estimates 
see \cite{Pfenniger} \cite{wdm_measurements_and_limits}:
equations (\ref{mh}) and (\ref{Th_Tg}) change by less than 1\%.
For a summary of measurements and their interpretation, see \cite{DM}.
For details of each measurement see the citations in \cite{comparing}.

\section{Adding particles to a self-gravitating isothermal gas}
\label{similar}

Consider a self-gravitating isothermal gas in equilibrium
with density $\rho(\mathbf{r})$.
We try to find a family of self-similar distributions by adding more particles
to the gas.
Distances scale as $\mathbf{r} \rightarrow \alpha \mathbf{r}$,
densities scale as $\rho(\mathbf{r}) \rightarrow \beta \rho(\alpha \mathbf{r})$,
and the particle velocities (including thermal velocities) scale as $v \rightarrow \gamma v$.
The temperature scales as $T \rightarrow \gamma^2 T$.
The gravitation field scales as $g(\mathbf{r}) = V^2/r \rightarrow (\gamma^2/\alpha) g(\alpha \mathbf{r})$ or
as $g(\mathbf{r}) = -G M/r^2 \rightarrow \alpha \beta g(\alpha \mathbf{r})$, and
the mass of the gas scales as 
$M(\mathbf{r}) \propto r^3 \rho \rightarrow \alpha^3 \beta M(\alpha \mathbf{r})$
or as $M(\mathbf{r}) \propto T r \rightarrow \gamma^2 \alpha M(\alpha \mathbf{r})$, so
\begin{equation}
\gamma = \alpha \beta^{1/2}.
\label{gamma}
\end{equation}
These relations are in agreement with equations (\ref{g})-(\ref{P2}).
The kinetic energy of the gas scales as $E_K \rightarrow \alpha^3 \beta \gamma^2 E_K$,
and the potential energy of the gas scales as $E_P \rightarrow \alpha^5 \beta^2 E_P$,
so the virial theorem $E_K = -E_P/2$ for the scaled gas is satisfied.

To complete the description of the problem at hand, we still
need to specify the equation of state of the gas, namely $T \rho(\alpha \mathbf{r})^{-2/3} =$ constant,
or $\gamma^2 = \beta^{2/3}$. We are then left with a
1-dimensional family of similar solutions with 
\begin{equation}
\gamma = \beta^{1/3} = \alpha^{-2}.
\label{alpha}
\end{equation}
All these similar solutions have the same adiabatic invariant $T \rho(\alpha \mathbf{r})^{-2/3}$.
During the accretion of new particles, 
the self-gravitating gas remains in thermal and mechanical equilibrium to a good approximation, and 
the adiabatic invariant remains constant.

\section{Galaxy formation}
\label{adding}

Consider a cored isothermal sphere.
According to (\ref{alpha}), this cored isothermal sphere is
defined by a single parameter, e.g. the core density $\rho_0$.
As the universe expands, new particles, if available, fall into the
halo potential well. 
The halo contracts as $\mathbf{r} \rightarrow \alpha \mathbf{r}$.
The converging particles increase the temperature of the halo
in proportion to $\gamma^2$, and the densities scale as 
$\rho(\mathbf{r}) \rightarrow \beta \rho(\alpha \mathbf{r})$,
where $\alpha$, $\beta$ and $\gamma$
satisfy (\ref{alpha}). To a good approximation, thermal and mechanical equilibrium
is maintained throughout this galaxy evolution.
The adiabatic invariant in the core of the galaxy remains constant,
i.e. $\gamma = \beta^{1/3}$.
 
As an example, assume that 
the core density $\rho_0$ of the cored isothermal sphere
increases by a factor 8. According
to the adiabatic invariant, 
or the noble gas equation of state,
the thermal velocity $\sqrt{\left< v_r^2 \right>}$
in the core increases by a factor 2, and
the temperature $T$ increases by a factor 4.
The Maxwell distribution of velocities in the core
depends on the ratio $v^2/T \rightarrow (2v)^2/(4T)$ that remains unchanged.
This Maxwell distribution of velocities in the core
implies that, in equilibrium, the halo is isothermal, so $T$ increases
by a factor 4, and $\sqrt{\left< v_r^2 \right>}$ increases by 
a factor 2,  both inside and outside of the core.
How is this possible?
In the present example, in the notation of equation (\ref{alpha}), 
$\beta = 8$, $\gamma = 2$, and $\alpha = 1/\sqrt{2}$.
So everywhere the gas is compressed, i.e. $r \rightarrow r/\sqrt{2}$,
$\rho(r) \rightarrow 8 \rho(r/\sqrt{2})$,
and $T \rightarrow 8 T$ everywhere.
The core radius $r_c$ decreases by a factor $2/\sqrt{8} = 1/\sqrt{2}$. 
The test particle rotation velocity $V$ increases by a factor 2. 
Note that the galaxy halo accretion of particles occurs in thermal and
mechanical equilibrium, i.e. adiabatically, to a good approximation.

We note that $r_\textrm{max}$ increases by a factor 2, 
and the mass $M(r_\textrm{max})$ of the halo
out to the new $r_\textrm{max}$ increases by a factor $\approx 8$.

The core density $\rho_0$ stops increasing, and the core radius $r_c$
stops decreasing, when there are no more particles entering the halo.
If the density peak in the early universe is surrounded by a negative
density fluctuation, the resulting galaxy will run out of new
particles, and will end up having a core.
If the density peak in the early universe is inside a positive 
density perturbation of larger size, the resulting galaxy
may end up with a core radius smaller than the observational
resolution, so this unresolved core is called a cusp. In conclusion, the core or
cusp of the galaxy depends on the available matter falling into
the galaxy halo. 
Note that it is not necessary to invoke dark matter self-interactions,
or baryonic feedback, to have a core.
%Note also that we have not considered black holes in this discussion.

To illustrate these comments, let us solve, by numerical integration, the following equations
that describe the formation of a galaxy:
Newton's equation
\begin{equation}
\nabla \cdot \mathbf{g} = -4 \pi G (\rho_h + \rho_b),
\label{Newton}
\end{equation}
the continuity equations
\begin{eqnarray}
\frac{\partial \rho_h}{\partial t}
& = & -\nabla \cdot (\mathbf{v}_h \rho_h), \label{continuity_h} \\
\frac{\partial \rho_b}{\partial t}
& = & -\nabla \cdot (\mathbf{v}_b \rho_b), \label{continuity_b}
\end{eqnarray}
and Euler's equations
\begin{eqnarray}
\frac{d \mathbf{v}_h}{d t} =
\frac{\partial \mathbf{v}_h}{\partial t} + (\mathbf{v}_h \cdot \nabla) \mathbf{v}_h
& = & (1 - \kappa_h(t)) \mathbf{g} - \frac{1}{\rho_h} \nabla (\left< v^2_{rh} \right> \rho_h), \label{Euler_h} \\
\frac{d \mathbf{v}_b}{d t} =
\frac{\partial \mathbf{v}_b}{\partial t} + (\mathbf{v}_b \cdot \nabla) \mathbf{v}_b
& = & (1 - \kappa_b(t)) \mathbf{g} - \frac{1}{\rho_b} \nabla (\left< v^2_{rb} \right> \rho_b). \label{Euler_b} 
\end{eqnarray}
$\mathbf{g}(\mathbf{r}, t)$ is the gravitation field, $\mathbf{r}$ is the proper 
(not comoving) coordinate vector,
$\rho_h(\mathbf{r}, t)$ and $\rho_b(\mathbf{r}, t)$ are the mass densities,
$\mathbf{v}_h(\mathbf{r}, t)$ and $\mathbf{v}_b(\mathbf{r}, t)$ are the velocity fields,
and $\sqrt{\left< v^2_{rh} \right>(t)}$ and $\sqrt{\left< v^2_{rb} \right>(t)}$
are the radial (1-dimensional) velocity dispersions, i.e. thermal velocities, 
of dark matter and baryons (mostly hydrogen), respectively.
Equations (\ref{Euler_h}) and (\ref{Euler_b}) express the conservation of momentum.
The static limits of equations (\ref{Newton})-(\ref{Euler_b}) are equations
(\ref{g})-(\ref{P2}). These equations need to be supplemented by the
equation of state of the gas.
The Maxwell distributions of baryons and dark matter 
have different temperatures $k T_b = m_p \left< v^2_{rb} \right>$ and
$k T_h = m_h \left< v^2_{rh} \right>$, respectively.

\begin{figure}
\begin{center}
%\vspace*{-4.5cm}
\scalebox{0.335}
%{\includegraphics{rho_190422.eps}}
{\includegraphics{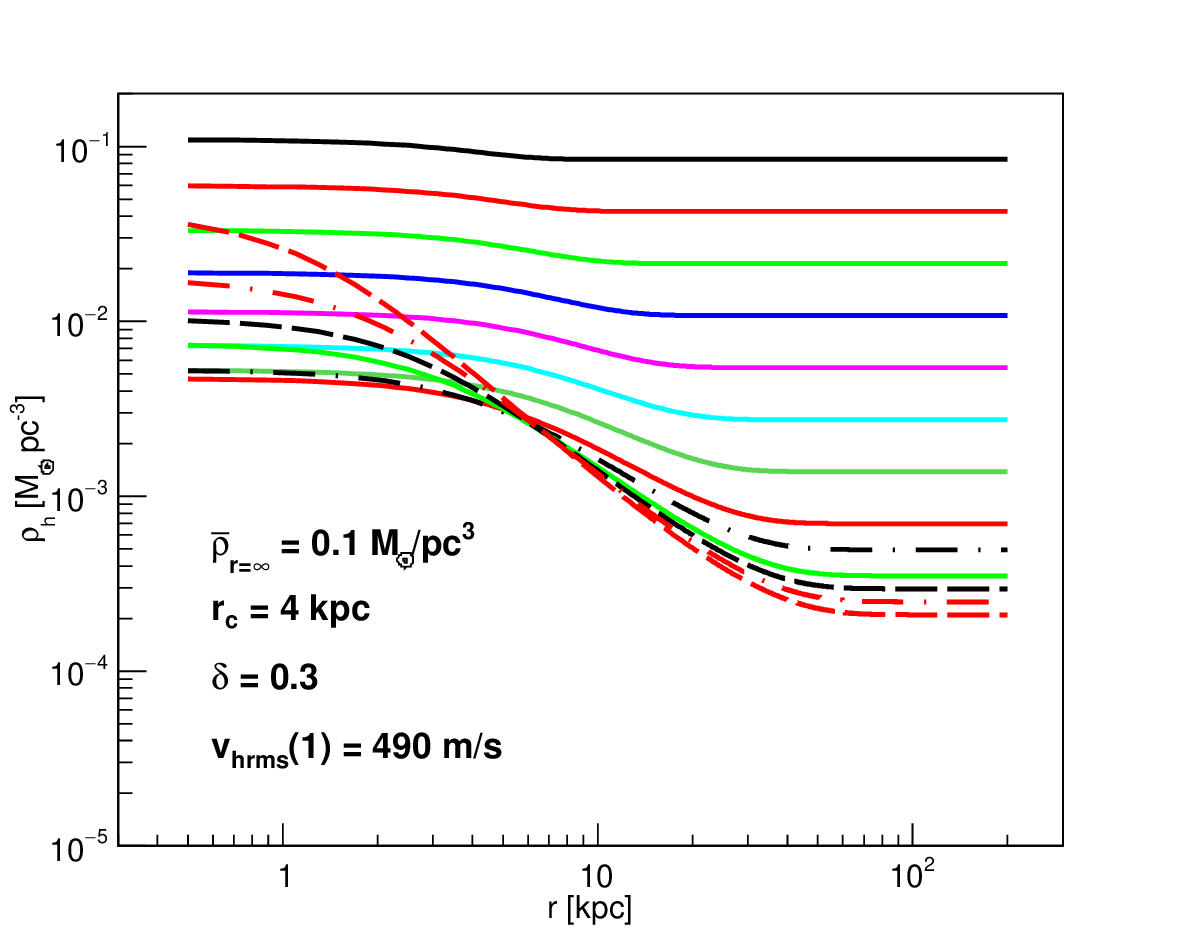}}
\scalebox{0.335}
%{\includegraphics{rhob_190422.eps}}
{\includegraphics{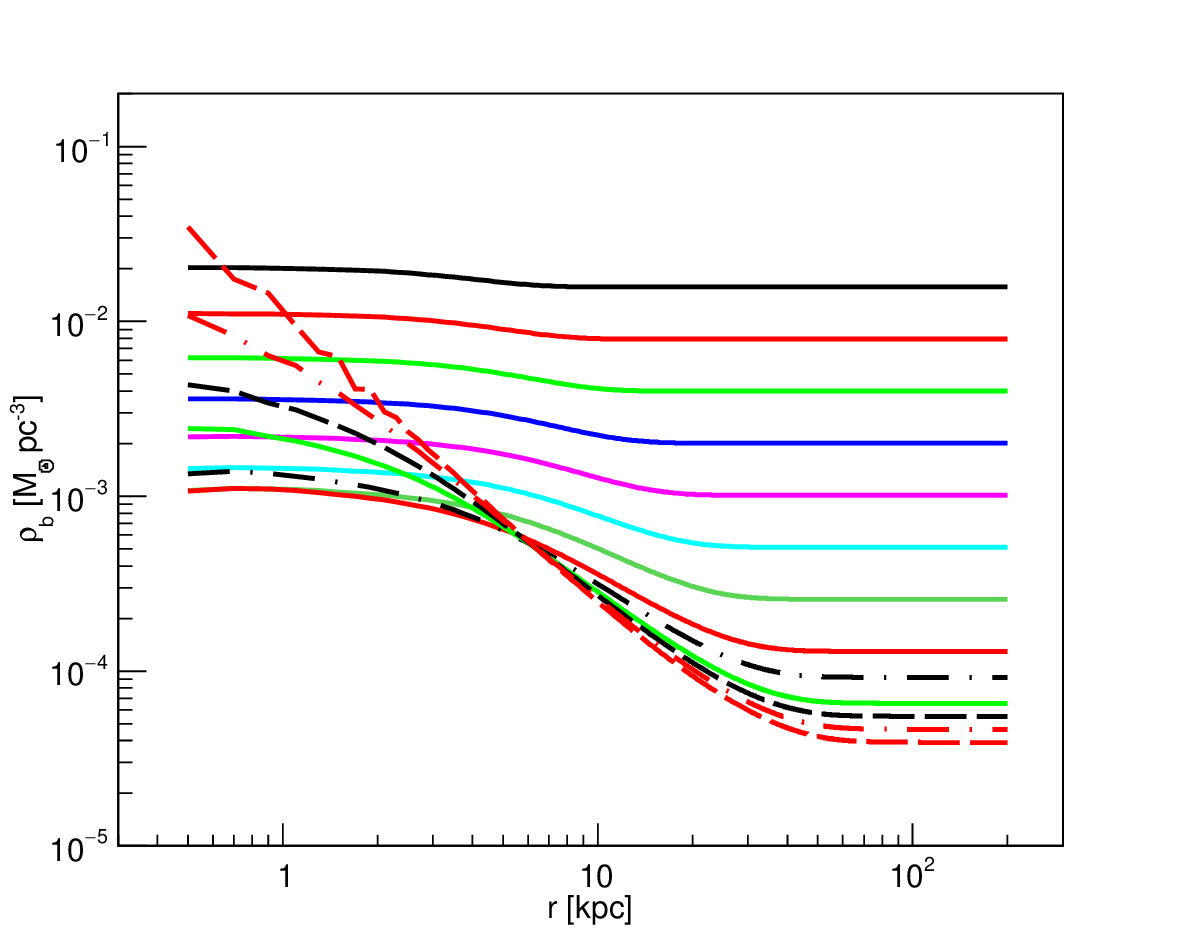}}
\scalebox{0.335}
%{\includegraphics{vr_190422.eps}}
{\includegraphics{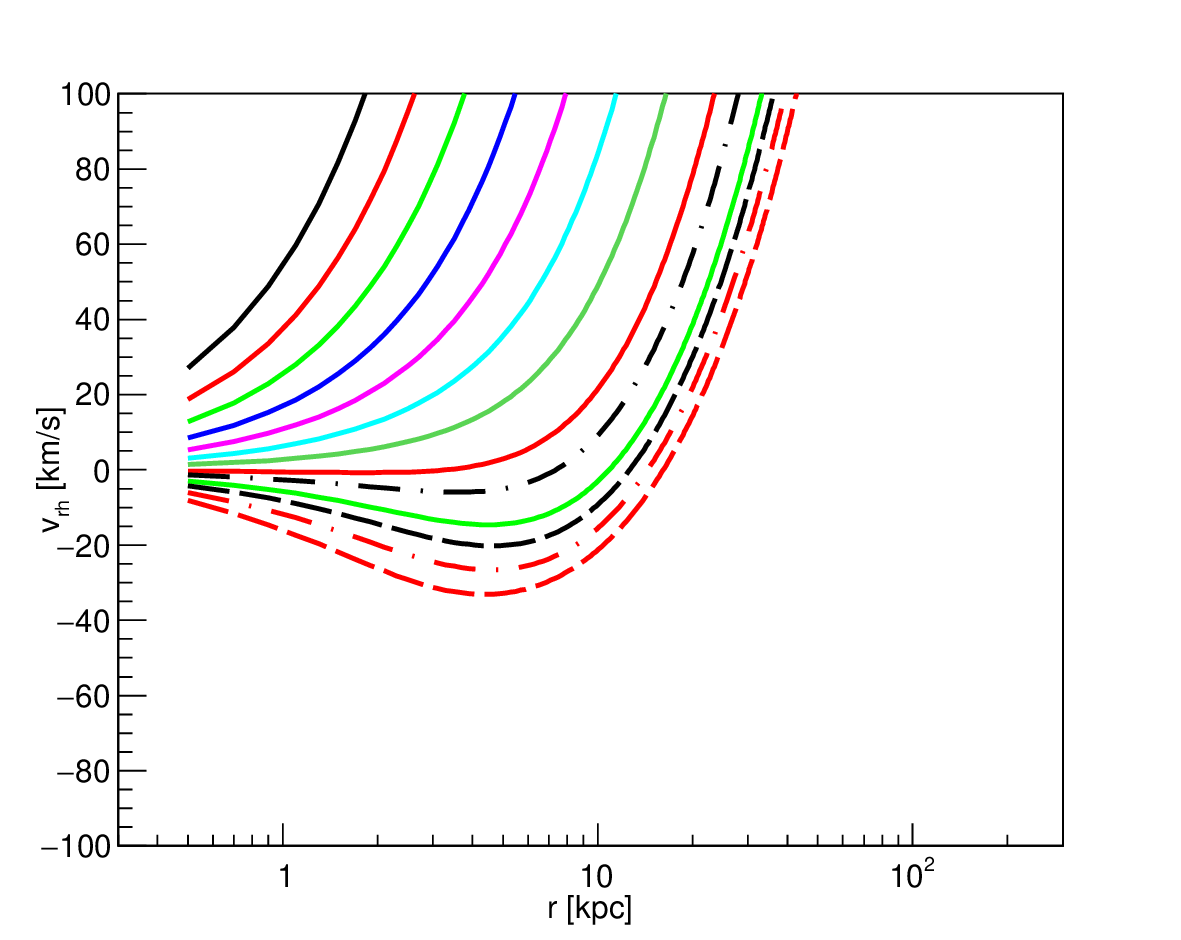}}
\scalebox{0.335}
%{\includegraphics{vrb_190422.eps}}
{\includegraphics{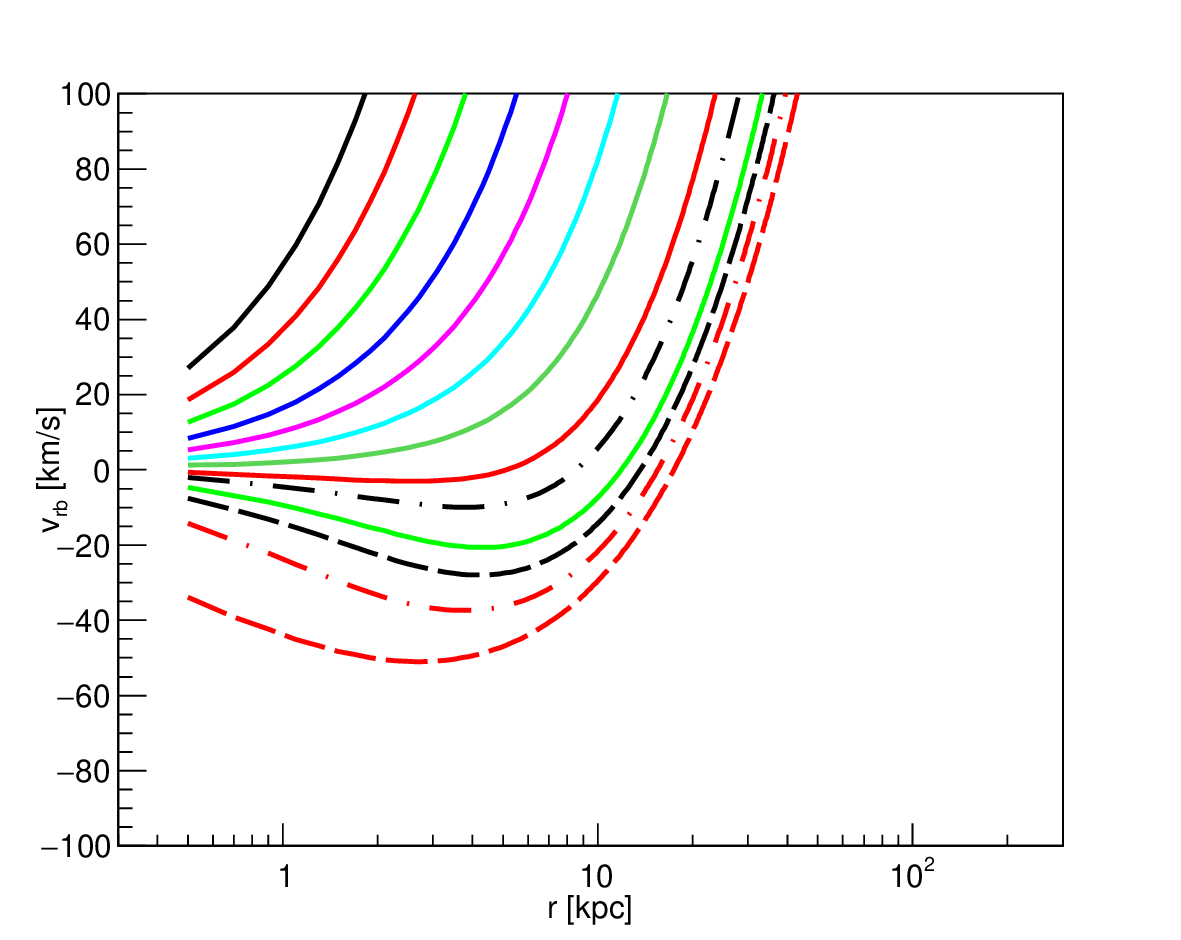}}
\caption{The formation of a warm dark matter plus baryon galaxy
with zero angular momentum is shown. The densities
$\rho_h(r)$ and $\rho_b(r)$, and the velocity fields $v_{rh}(r)$ and $v_{rb}(r)$,
are presented at time-steps that increase by factors 1.4086
(or $\sqrt{1.4086}$ for the dot-dashed lines).
The initial perturbation is Gaussian, with parameters listed in the figure.
Dark matter is warm with $v_{h\textrm{rms}}(1) = 490$ m/s.
}
%galaxy_h_b_adiabatic.C_bck011023
%/home/bruce1/first_star/first_stars.tex
%galaxy_h_b.C_bck260222_3
%galaxy_h_b.C_bck270222
%galaxy_h_b.C_bck240322_2
%galaxy_h_b.C_bck140422_2
%galaxy_h_b.C_bck180422_3
%galaxy_h_b.C_bck190422
\label{galaxy_h_b_12}
\end{center}
\end{figure}

As an example we assume no rotation, i.e. $\kappa_h(t) = \kappa_b(t) = 0$,
and set the adiabatic invariant to $v_{h\textrm{rms}}(1)=490$ m/s.
For baryons we take $v_{b\textrm{rms}}(1) = 21$ m/s, corresponding
to hydrogen decoupling from photons at $z \approx 150$ \cite{Weinberg}.
We set the initial $\rho_h(r)$ and $\rho_b(r)$ as shown in Figure \ref{galaxy_h_b_12}.
We integrate the equations in steps $dt$, and for each $t$, in steps $dr$,
starting at $r_\textrm{min}$, to calculate
the new $\rho_b(r, t+dt)$ and $\rho_h(r, t+dt)$.
The above equations are supplemented by the adiabatic conditions, so,
for each step of $t$, we set, only once, i.e. for all $r$:
\begin{eqnarray}
\sqrt{\left< v_{rh}^2 \right>(t)} & = & \frac{v_{h\textrm{rms}}(1)}{\sqrt{3}}
\left( \frac{\rho_h(r_{\textrm{min}}, t)}{\Omega_c \rho_{\textrm{crit}}} \right)^{1/3}, \qquad \textrm{and} \\
\label{vhrms2}
\sqrt{\left< v_{rb}^2 \right>(t)} & = & \frac{v_{b\textrm{rms}}(1)}{\sqrt{3}}
\left( \frac{\rho_b(r_{\textrm{min}}, t)}{\Omega_b \rho_{\textrm{crit}}} \right)^{1/3}.
\label{vbrms2}
\end{eqnarray}
This isothermal prescription, valid at	least out to the last observed rotation	velocity,
can not be correct beyond $r_\textrm{max}$
where the universe is expanding homogeneously. However, this is not a problem
since, beyond $r_\textrm{max}$, $\nabla (\left< v^2_{rh} \right> \rho_h)$
and $\nabla (\left< v^2_{rb} \right> \rho_b)$ are zero.
While the hydrogen and helium gas remains adiabatic,
i.e. until excitations, radiation, shocks and star formation become significant, we require
(\ref{vbrms2}) in the core of the galaxy. 
An example of the formation of a galaxy is shown in Figure \ref{galaxy_h_b_12}.

\section{Conclusions}

We have studied galaxies, and galaxy formation, assuming that dark matter is warm.
This scenario requires the addition of one parameter to the 
cold dark matter $\Lambda$CDM
cosmology, namely the adiabatic invariant $v_{h\textrm{rms}}(1)$.
We find that the formation of galaxies, all the way from linear perturbations in the
early universe, until the galaxies run out of new particles to accrete, is adiabatic
to a good approximation, and that the adiabatic invariant in the core of the galaxy is conserved. 
The run out of new matter to accrete determines the final density and radius of the galaxy core
(note that it is not necessary to invoke dark matter self interactions or baryonic feedback
to justify a core).
The observed spiral galaxy rotation
curves allow measurements of $v_{h\textrm{rms}}(1)$ \cite{dwarf}. 
These measurements are consistent for galaxies with absolute luminosities
spanning $3.5$ orders of magnitude \cite{dwarf}, so the analysis
is validated by observations, and the interpretation that $v_{h\textrm{rms}}(1)$
is of cosmological origin is confirmed.
Independent measurements
of $v_{h\textrm{rms}}(1)$ are obtained by studying the consequences of the
warm dark matter free-streaming suppression factor $\tau^2(k)$ of the
cold dark matter power spectrum of density perturbations, i.e.
galaxy stellar mass distributions,
galaxy rest-frame ultra-violet luminosity distributions,
first galaxies,
and their effect on the reionization optical depth \cite{UVL}.
All of these measurements of $v_{h\textrm{rms}}(1)$ are consistent,
and constrain the warm dark matter particle properties
%, i.e. mass, temperature, spin and couplings 
\cite{DM} \cite{comparing} \cite{inflation}.


\begin{thebibliography}{7}

\bibitem{PDG2020} %1
Particle Data Group,
Zyla, P.A., \textit{et al.}
(2020) Review of Particle Physics,
\textit{Progress of Theoretical and Experimental Physics} \textbf{2020}, 083C01.

\bibitem{Pfenniger} %2
Paduroiu, S., Revaz, Y., Pfenniger, D. (2015)
Structure formation in warm dark matter cosmologies Top-Bottom Upside-Down.
https://arxiv.org/pdf/1506.03789.pdf

\bibitem{SPARC} %3
Lelli F., McGaugh S. S., Schombert (2016),
SPARC: Mass models for 175 disk galaxies with Spitzer Photometry and Accurate Rotation Curves
\textit{The Astronomical Journal}, 152:157.
doi:10.3847/0004-6256/152/6/157.
The data in digital form is publicly available in files
\textsf{SPARC\_Lelli2016c.mrt} and \textsf{LTG\_data.txt}.

\bibitem{part1}  %4
Hoeneisen, B.
(2019) A Study of Dark Matter with Spiral
Galaxy Rotation Curves.
\textit{International Journal of Astronomy and Astrophysics}, 9, 71-96.
%https://www.scirp.org/pdf/IJAA_2019043016375354.pdf

\bibitem{dwarf} %5
Hoeneisen, B. (2022) Measurement of the Dark Matter Velocity Dispersion with
Dwarf Galaxy Rotation Curves,
\textit{International   Journal   of   Astronomy   and   Astrophysics}, \textbf{12}, 363-381.
%/home/bruce1/THINGS/THINGS_ijaa_190922.tex

\bibitem{comparing}  %6
Hoeneisen, B. (2023) Comparing measurements and limits on the
warm dark matter temperature-to-mass ratio.
https://arxiv.org/pdf/2308.10356.pdf

\bibitem{wdm_measurements_and_limits} %7
Hoeneisen, B. (2022)
Comments on Warm Dark Matter Measurements and Limits.
\textit{International  Journal  of  Astronomy  and Astrophysics}, \textbf{12}, 94-109.

\bibitem{DM} %8
Hoeneisen, B. (2023) A Data Driven Solution to the Dark Matter Problem.
\textit{European Journal of Applied Sciences} \textbf{11}, 473-481.
%https://drive.google.com/file/d/1tyNR1BRcxbxaqbeSoj0kSxQ7-uoHDxNQ/view

\bibitem{Weinberg} %9
Weinberg, S. (2008) Cosmology. Oxford University Press.

\bibitem{UVL} %10
Hoeneisen, B. (2022) Measurement  of  the  Dark  Matter  Velocity   Dispersion   with   Galaxy
Stellar   Masses, UV Luminosities, and Reionization.
\textit{International   Journal   of   Astronomy   and   Astrophysics}, \textbf{12}, 258-272.

\bibitem{inflation} %11
Hoeneisen, B. (2023) Exploring Inflation Options for Warm Dark
Matter Coupled to the Higgs Boson,
\textit{International   Journal   of   Astronomy   and   Astrophysics}, \textbf{13}, 217-235.

\end{thebibliography}
\end{document}